# Beyond time delays: How web scraping distorts measures of online news consumption

Roberto Ulloa(*)[1,2], Frank Mangold[1], Felix Schmidt[1], Judith Gilsbach[1,2], Sebastian Stier[1,3]

[1] GESIS – Leibniz Institute for the Social Sciences, Cologne
[2] University of Konstanz
[3] University of Mannheim

(*) Corresponding author: roberto.ur@pm.me, roberto.ulloa@gesis.org, roberto.ulloa@uni-konstanz.de

# Abstract

As the exploration of digital behavioral data revolutionizes communication research, understanding the nuances of data collection methodologies becomes increasingly pertinent. This study focuses on one prominent data collection approach, web scraping, and more specifically, its application in the growing field of research relying on web browsing data. We investigate discrepancies between content obtained directly during user interaction with a website (in-situ) and content scraped using the URLs of participants' logged visits (ex-situ) with various time delays (0, 30, 60, and 90 days). We find substantial disparities between the methodologies, uncovering that errors are not uniformly distributed across news categories regardless of classification method (domain, URL, or content analysis). These biases compromise the precision of measurements used in existing literature. The ex-situ collection environment is the primary source of the discrepancies (~33.8%), while the time delays in the scraping process play a smaller role (adding ~6.5 percentage points in 90 days). Our research emphasizes the need for data collection methods that capture web content directly in the user's environment. However, acknowledging its complexities, we further explore strategies to mitigate biases in web-scraped browsing histories, offering recommendations for researchers who rely on this method and laying the groundwork for developing error-correction frameworks.

# Introduction

Digitalization presents unprecedented methodological challenges in measuring media use (Andersen et al., 2016), as citizens' information exposure has become increasingly fragmented (Messing & Westwood, 2014; Möller et al., 2020). Accordingly, scholars have turned to computational methods in media and communication research as they contribute new forms of data to address the validity and reliability issues inherent in traditional survey self-report measures (Parry et al., 2021; Scharkow, 2016). One approach that has gained traction in recent years is web-tracking (Bachl et al., 2024; Guess, 2021; Scharkow et al., 2020; Stier et al., 2022; Ulloa & Kacperski, 2023) as it holds the promise of overcoming the pitfalls of previous studies that inferred the contents media users were exposed to from survey self-reports (Scharkow & Bachl, 2017) by favoring direct observations of such exposure. Web-tracking allows for a more precise measurement of various phenomena, such as audience fragmentation and polarization, selective exposure, motivated reasoning, and media effects on attitudes and behavior. Still, previous web-tracking studies have predominantly relied on data from commercial market research institutes or data donations that only included the URLs of participants' website visits. In essence, the data did not allow for first-hand measures of the content participants were exposed to, but scholars had to indirectly approximate the content through an ex-situ, post-hoc web scraping of the URLs visited by participants.

Theoretically, we build and expand on previous literature to argue that ex-situ post-hoc web scraping can compromise web-tracking results of content exposure by introducing sampling bias (Foerderer, 2023), leading to measurement errors in content analysis and smaller effect estimates (Scharkow & Bachl, 2017). First, websites and their content change over time and cannot always be accessed retroactively; e.g., Dahlke et al. (2023) report systematic differences caused by scraping data at different points in time. Second, a website's content varies depending on the user environment, for example, their web page sessions and the region from which it is accessed (Foerderer, 2023). In contrast, in-situ web-tracking methodologies (e.g., Adam et al., 2024) gather browsing content on the individuals' environment in real-time, offering high-fidelity portrayals of user experiences that allow behavioral studies of fine granularity (e.g., Schmidt et al., 2024; Ulloa & Kacperski, 2023).

Against this theoretical and methodological background, we conceptualize and empirically quantify the biases that web scraping introduces to web-tracking research that relies on measures of news content exposure. Furthermore, we establish and demonstrate how recently developed academic alternatives to commercial web-tracking data help to alleviate these biases by directly using them in new studies and – more generally – by laying the groundwork for developing error-correction frameworks. Our study relies on a comprehensive form of gathering browsing data that directly collects the content during the user interaction in the user's environment (in situ). Equipped with these data, we first measure the information disparity resulting from a stringent scenario, namely, web-scraping the content in *near real-time* from a dedicated server infrastructure (ex-situ). Second, we study the detrimental effect of post-hoc scraping by systematically quantifying the error against the actual content presented to the participant, going beyond previous measurements limited by ex-situ collections (Dahlke et al., 2023). To probe the precision and time-sensitivity of web-scraping approaches, we repeat our post-hoc scraping with delays (30/60/90 days) calculated exactly after each user interaction's timestamp. Third, we explore diverse strategies to mitigate the biases occurring in automatic content classification applied to the scraped content, shedding light on the roots of the problem.

We focus on web pages that correspond to news articles. First, much of existing web-tracking research is related to online news exposure (Scharkow et al., 2020; Stier et al., 2020), including studies that have scraped the content to uncover the information the visitors of news sites get exposed to (Flaxman et al., 2016; Guess, 2021; Jürgens & Stark, 2022; Reiss, 2023; Stier et al.,

2022). Second, news articles are expected to be rarely deleted or change in content as they should undergo a more careful review process under journalistic guidelines before publication, i.e., their content is stable, especially when compared to other web pages, such as news homepages or category indexes (web pages that list news under a particular category, e.g., politics), search engines, or social media. Methodologically, they serve as a conservative case study as news articles are less likely to suffer the biases we quantify in this paper. Third, news articles are also more likely to be investigated as they serve as a strong indicator of exposure and are thematically invariant (the content of one news article is usually germane to one topic only); hence, they constitute a more suitable unit of analysis, e.g., for automatic classification in machine learning (Barberá et al., 2021; Boumans & Trilling, 2016). Given the advances in natural language processing that simplify content annotation (Gilardi et al., 2023; Patadia et al., 2021), it is likely that the practice will become increasingly common.

Our results demonstrate stark disparities between in-situ and (near real-time) ex-situ data collections; as a lower-bound estimate, at least 33.8% of the contents of participants' online news exposure cannot be determined using static web scraping approaches. While most widely accentuated in previous literature, the issue of temporal delays (post-hoc) was of relatively minor importance; a 30-day delay in the scraping will increase the above disparity by ~5.55 percentage points, and a 90-day delay would only increase it further by less than one percentage point. Furthermore, we contribute evidence that such disparities are not uniformly distributed and compromise the validity of findings by introducing systematic biases. Finally, we discuss the reasons behind the disparities and offer recommendations for future works that use web-scraped browsing histories, emphasizing the need to develop a robust method for identifying pages that require user interaction (e.g., paywalls) as the most critical source of content distortion.

**The challenge of measuring online news exposure**

Despite being crucial to understanding media effects, measuring media exposure remains challenging in communication research (Andersen et al., 2016; Bartels, 1993). The shift to an online environment in the last decades has complicated the challenge by increasing the variety of platforms and information sources individuals can choose from (Van Aelst et al., 2017; Vliegenthart, 2022), leading to a paradigm wherein individuals actively select the information they consume according to their preferences (Bennett & Iyengar, 2008; Chaffee & Metzger, 2001; Schulz, 2004; Tewksbury, 2003). Meanwhile, this transition has introduced new data collection methods (Christner et al., 2022; Ohme et al., 2023), offering opportunities to address limitations associated with self-reporting by observing data directly (Parry et al., 2021; Scharkow, 2016). Indeed, media and communication scholars could traditionally only infer the contents media users get exposed to by combining survey self-report measures of media use from surveys with content analytical measures of the messages published by media outlets – an approach that has arguably contributed to understanding media effects yet been restricted by measurement error in both survey and content data (for reviews, see De Vreese et al., 2017; Scharkow & Bachl, 2017).

Web-tracking is advancing the field by offering a more reliable estimation of media exposure and its effects (Parry et al., 2021; Scharkow, 2016), with an increasing number of studies exploiting the method. For example, research explores the role of online intermediaries (e.g., social media and search engines), increasing news exposure and diversity (Scharkow et al., 2020; Stier et al., 2022), the role of search engine ranking driving diverse audiences to outlets and users consuming news from different sources (Ulloa & Kacperski, 2023), the prevalence of health-related search engine use (Bachl et al., 2024), and the constrained existence of extreme ideological news exposure chambers only among a few individuals (Guess, 2021). Nevertheless, even with observable data, the measurement of media effects on attitudes and polarization remains challenging, with recent experimental evidence finding small or null effects (Allcott et al., 2020; Casas et al., 2023; Guess, 2021; Levy, 2021; Wojcieszak et al., 2022), which might also be due to issues associated with measurement (Scharkow & Bachl, 2017).

Many such studies have relied on aggregating visits to news sources, which offers a narrow picture of media exposure. Conversely, studying social phenomena such as news audience fragmentation and polarization necessitates precise accounts of political news exposure, including granular measures that adequately capture the relevant dimensions of user interactions (Tyler et al., 2022; Yang et al., 2020). Conceptually, the qualification of news based on sources is limited, and instead, the measures should focus on analyzing the content published by them (Fletcher & Nielsen, 2017; Kroon et al., 2024). First, online users engage with news content while often unaware of the sources of the articles they engage with (Kalogeropoulos et al., 2019; Messing & Westwood, 2014). Second, various other concepts related to media and communication research, like negativity (Atteveldt et al., 2021), misinformation (Aslett et al., 2024), agenda setting (Burscher et al., 2015), and framing (Budak et al., 2016) can only be feasibly studied based on knowledge of the contents of media exposure. Third, the analysis based on sources relies on strong assumptions regarding the consistency of the content with the source's values (Ganguly et al., 2020; Tyler et al., 2022). In general, the reliance on sources leads to measurement error, i.e., misestimations of the information distribution, that would perturb the estimation of media effects.

Nevertheless, gaining access to the actual content behind the sources that individuals visit is not trivial. Standard solutions for web-tracking, e.g., commercial tools and data donations of browsing histories, only provide metadata regarding the visit. The metadata usually includes the visited URL (Uniform Resource Locator) used to identify and access the content behind the visit. Researchers have used web-scraping to complement their collection and subsequently classify the gathered content, e.g., to identify news (Flaxman et al., 2016; Reiss, 2023), opinion pieces (Flaxman et al., 2016), and political news (Guess, 2021; Stier et al., 2022).

Additionally, although web-scraping is conceptually a simple process, in practice, numerous considerations have received little attention in communication research besides discussing technical and legal hurdles to confront (Freelon, 2018). The considerations are germane to the difference in which the environment is accessed: either in the individual's device and browsing configuration (in-situ) or the researcher's infrastructure (ex-situ). First, there are apparent caveats regarding the limitations of accessing social media data, as the news post feeds (aka timelines or news feeds) entirely depend on the logged user, and there are numerous ways in which the visibility of the message is restricted that require alternative approaches (Adam et al., 2024; Haim & Nienierza, 2019). Second, the dynamism of the online environment demands the quick gathering of the content before it changes. Dahlke et al. (2023) studied the impact of time delays in the accessibility of news pages, e.g., due to server errors (e.g., such as web pages removal, i.e., 404 errors) or due to the use of paywalls or other restriction mechanisms (for a review on the evolution of news payment systems, see Arrese, 2016). They found that time delays in scraping negatively affect the unrestricted retrieval of "hard news" web pages, while the retrieval of misinformation pages is relatively stable. They found a modest time effect (2.2% in one year); however, their first scraping snapshot was taken one year after the user visited the web pages and used it as a baseline for their measurement. Third, in their analysis, there is an implicit assumption that individuals are not similarly affected by such restrictions, which is certainly not the case (Olsen et al., 2020). Fourth, while Dahlke et al. (2023) shed light on the evolution of restricted content (such as login pages or paywalls and server errors, e.g., 404 errors), there are other sources of variation, such as advertising, personalized or time-sensitive recommendations, and direct text modifications (e.g., error corrections) that have not been accounted for. The last three considerations affect the content of news information pieces.

Although recent advances make it possible to correct biases introduced by misclassification in automated content analysis (TeBlunthuis et al., 2024), they indirectly rely on high-quality content as they acknowledge the assumption of high-quality human annotations. For example, Dahlke et al. (2023) found that 11.6% of their manually annotated visits were categorized as having restricted content, and thus, the human annotators would be limited in their assessment. Even if the restrictions are circumvented in the web-scraping process, it is unclear if the individuals were also blocked on those pages. Confined within this entangling scenario, researchers have developed increasingly comprehensive web-tracking solutions that aim to capture the content as seen by the user (Adam et al., 2024). Equipped with this data, we quantify and explore the gap between collecting web data directly in the user environment (in-situ) and collecting it via scraping in remote servers (ex-situ), including a fairer quantification of time effects.

## Research Questions

There are two primary alternatives to capture the content (HTML) that participants consume: in-situ collections capture content in the user environment (which includes their device and browser, but also aspects such as if the user is logged in, cookies, and sessions), while ex-situ collections use an external server to scrape the content from the URLs that the participants had visited from. The in-situ method happens in real-time, while the ex-situ collection happens necessarily after the interaction. The main goal of this study is to quantify the discrepancy between the high-fidelity in-situ collection and the ex-situ collection:

> ***RQ1: What is the disparity between the content of the pages collected via in-situ and near real-time ex-situ methods?***

We compare the performance of ex-situ methods against our robust in-situ baseline across three representations of the scraped content: (1) cleaned text, targeting the core of the news content by removing boilerplate (e.g., menus and ads) but that potentially introduces noise (due to extraction errors); (2) raw text, that includes boilerplate but is overall more reliable (no extraction errors); and the (3) full content (i.e., plain HTMLs).

We hypothesize that (H1a) such discrepancies will be higher than 0. We compare the content using the (Levenshtein) distance between three representations of the content. Due to the amount of personalized content included in the raw text (e.g., advertising) and the larger amount of boilerplate information present in the HTML (e.g., user metadata), we hypothesize that (H1b) the discrepancies on the cleaned text will be smaller than those on raw text and (H1c) those on raw text smaller than with the plain HTML. To improve the accuracy of our estimates, we repeated our experiment in two waves. Given that the waves included different participants generating the data at various times, we expect differences between the two waves (H1d). To partially account for potential differences in the operationalization of the ex-situ collection, we included two commonly used libraries to scrape content (for details on the libraries and argumentation of the selection, see Data collection): we hypothesize that (H1e) the discrepancies will differ between the two scraping libraries due to different parameterizations of the libraries, e.g., the user agent (a signature of the user environment or scraper).

Considering the dynamic nature of the web, a delay in the post-hoc ex-situ collection may lead to additional discrepancies. Thus, we also assess data disparity at 0, 30, 60, and 90 days following the in-situ captures:

> ***RQ2: Does delaying the ex-situ data collection exacerbate the data disparity compared to in-situ collections?***

In line with previous research (Dahlke et al., 2023), we hypothesize that (H2) the longer the delay is, the larger the discrepancy between the ex-situ and in-situ collections.

News outlets have different policies regarding access to their online content. For example, they implement different revenue models that result in differences in the prevalence of paywalls (Arrese, 2016; Myllylahti, 2017; Sjøvaag, 2016); these revenue models are associated with the type of news outlets (e.g., commercial broadcasters as opposed to legacy press) or the type of content covered (political or entertainment). In general, such policies may affect the ex-situ data collection disproportionally across different types of news websites.

> ***RQ3: Are the discrepancies between content obtained via ex-situ and in-situ data collections uniformly distributed across news types?***

We formulate hypotheses based on three different news categorizations. The first categorization centers on well-known major variations across content producers in the German media system. Due to different degrees of professionalization, we expect that (H3a) discrepancies in the content scraping are not uniformly distributed depending on whether a news outlet is, e.g., a public broadcaster, legacy press outlet, or digital-born outlet (see Methods; Stier et al., 2020). We also hypothesize that (H3b) the discrepancies are not uniformly distributed according to the internal categories assigned by the news outlets through keywords in the URL path (e.g., example.com/politics/ would suggest the politics category; see Methods). Finally, there might also be (H3c) discrepancies between the ex-situ and in-situ collections in the distributions of pages to categories according to a content classification operationalized via a machine learning (ML) classifier (Erfort et al., 2023).

Given the observed differences between in-situ and ex-situ data collections across various news content types, we explore strategies to mitigate biases inherent to the ex-situ collection approach.

> **RQ4: Can specific strategies minimize discrepancies between content obtained via ex-situ and in-situ data collections?**

To address this research question, we will focus on the systematic biases found in content-based categorization (Erfort et al., 2023). Specifically, we will examine the efficacy of (1) adjusting the classifier's (score) thresholds, (2) excluding the most affected news domains, and (3) identifying the sources of the bias and excluding the most affected categories. To operationalize RQ4, we will compare the distribution of news categories between ex-situ and in-situ collections; the closer the distribution is, the better the strategy.

# Methods

**Participants**

Participants were recruited from GESIS (Leibniz Institute for the Social Sciences)'s new panel whose members are recruited via Meta ads (Instagram and Facebook) and ALLBUS, the German general social survey representative of the adult population in Germany. Individuals were invited to install a desktop Chrome/Firefox/Edge browser extension (Adam et al., 2024; Aigenseer et al., 2019), which collects data about website visits from participants' web browsers. The data includes the domain, full URL, content (HTML), and timestamps of visits. The major advantage of the browser extension is the direct gathering of content seen by participants. These data collected in-situ will serve as a benchmark for assessing the performance of ex-situ data collection approaches.

Recruitment occurred in two waves comprising 534 individuals: 227 participated in the first wave from July 5th to October 9th, 2023 (637 persons were invited in a recruitment survey from July 5th to August 9th). Participants could earn a compensation of up to 5 euros of unconditional prepaid incentives and up to 40 euros of postpaid incentives conditional on at least 30 days of activity within a study period of 60 days. The incentive payment was embedded in a recruitment experiment that was running in parallel; the maximum compensation was, therefore, 45 euros.

493 persons participated in the second wave from October 16th to December 7th, 2023 (1,772 persons invited; same compensation scheme). Exclusions in the recruitment into the web tracking occurred for individuals that (a) were not German residents, (b) did not use any of the supported browsers (Firefox, Chrome, or Edge) as their primary one, (c) did not use a desktop computer, or (d) did not consent to the data collection or not install the extension despite giving consent. During both waves, ~2.9 M page visits of 534 participants were recorded. We analyzed 34,108 visits to news articles (see Data filtering) of 412 participants: 12,859 visits by 180 participants for wave 1, 22,321 visits by 364 participants for wave 2, and 12,959 visits by 315 participants for the last 30 days of wave 2 (i.e., the subsample used for the post-hoc experiment). 131 individuals participated in the two waves. See Appendix S1 for other statistics on web activity and participants' demographic characteristics by wave.

**Data protection**

Web-tracking data collection was approved by the GESIS ethics committee (Decision-2021-6) and conducted in line with the institute's data protection and ethics regulations, including collecting explicit informed consent to install a plugin in their primarily used desktop browser. The plugin avoided collecting sensitive websites or information, such as bank or insurance content, or third-party individuals (e.g., social media or emails), using pre-generated deny lists and an external API service (webshrinker.com). Participants could temporarily deactivate the tracking at any time. Anonymized data sets can be shared to reproduce the presented analysis and plots upon request.

**Data collection**

Given a participant's visit to a webpage, the browser extension collected the URL, content (HTML), and the visit's landing time. This information (in-situ) was sent to our collection servers once the participant left the page (either closed the page or navigated to another page in the same tab). Once the external servers received the information, the URL was randomly assigned to one of two Python scraping libraries: trafilatura v1.6.1 (Barbaresi, 2021) or requests 2.27.1 (Reitz, 2023). This randomization avoided the duplication of requests from our collection server; news outlet servers would have been more likely to block us if they received a higher volume of duplicated traffic from our server, thus inflating our estimates. We parallized the scraping processes to avoid further delays, especially during peak hours.

The two libraries were selected due to their popularity and prominence in the research community. According to Barbaresi (2021), trafilatura is the most accurate library for extracting clean text from HTML; we used its pre-configured scraping functionality, which includes a predefined user agent of the HTTP request (i.e., a self-identified signature, which either describes the user environment, e.g., the browser, or whether the agent corresponds to a scraping bot). Due to its effectiveness in extracting text, trafilatura is commonly used in academic contexts. The requests library is likely the most used Python library for web scraping and is highly configurable; for this study, we set the request headers to block all cookies, allow redirects and timeout after 30 seconds and randomized the user agent sampling from a pool of common ones (Wittman, 2019/2022). Given the stringent conditions of our near real-time experiment involving participants, we did not include tools that facilitate the dynamic loading of content (e.g., Selenium, which runs a browser without a graphical interface in the background), usually via automatic scrolling. Such tools would have raised the requirements of our data collection infrastructure beyond our capabilities (e.g., running and monitoring heavy parallel processes). We argue that Selenium would not substantially affect the results reported in our study because the content of news articles is primarily static, i.e., they do not require dynamic loading more commonly used for web elements such as social media feeds.

The content is scraped immediately after the library is assigned (near real-time ex-situ). Additionally, the URLs collected in the last 30 days of the in-situ collection were queued for 30/60/90 days and scraped accordingly (post-hoc ex-situ) with the assigned library. We only take the last 30 days of the in-situ collection to avoid two concurrent post-hoc collections, which would contaminate the observations by putting more pressure on external servers. For the 30-day post-hoc collection, each visit is delayed by 30 days; thus, the post-collection takes as much time as the in-situ collection, finishing just in time for the start of the 60-day post-hoc collection, and so on. There was an extra delay on less than ~1.5K pages due to a technical error from 19.02.2024 to 22.02.2024 (i.e., the most affected page was delayed ~3.5 days).

To help disentangle the potential sources of discrepancies between the in-situ and ex-situ collections, we present the number of HyperText Transfer Protocol (HTTP) errors according to scraping libraries and each ex-situ subcollection (Table 1). As the underlying internet protocol, HTTP describes the rules for transferring web pages. HTTP errors indicate issues encountered while accessing web pages, ranging from missing pages to connectivity problems to server failures. The low number of errors in the table indicates a good overall technical performance of the scraping procedure, especially if we look at Wave 1 (i.e. before external servers would have detected abnormal activity). These errors contribute to the discrepancies between the collections and are, thus, always included in our study (the content is assumed to be empty). We can see that the number of errors increases in the latter collections, albeit, e.g., the scraped pages (and corresponding script) of the post-hoc experiment are the same. Moreover, relying entirely on HTTP errors would be wrong because different external (new outlet) servers can take various actions when, e.g., a removed page is deleted. The server can be configured to send an HTTP error with no content or send a valid page indicating that such a page does not exist.

| Collection Type | Requests Failures | Trafilatura Failures | Total Failures | N |
|---|---|---|---|---|
| Wave 1 | 4 | 67 | 35 | 12326 |
| Wave 2 | 31 | 619 | 686 | 21782 |
| 0-day post-hoc | 35 | 369 | 404 | 12668 |
| 30-day post-hoc | 214 | 490 | 704 | 12668 |
| 60-day post-hoc | 225 | 503 | 728 | 12668 |
| 90-day post-hoc | 237 | 514 | 751 | 12668 |

**Table 1. HTTP errors in each of the ex-situ collections.** The first column indicates the collections, followed by the number of errors according to each library and the total of errors. The last column shows the total number of articles scraped and used in the study. Note that the 0-day post-hoc is a subset of Wave 2, which is scraped again in the 30/60/90 days post-hoc collections.

**Data filtering**

We only considered URLs belonging to German news domains identified using a list of 679 domains provided by Stier et al. (2020), excluding homepages identified using a domain match with the news domain list and heuristics to identify patterns, e.g., URLs that end just after the domain (e.g., [https://example.com/](https://example.com/)) or that include only keywords (e.g., index, home) in the path of the domain (e.g., [https://example.com/index.html](https://example.com/index.html)). Once the data collection was finished, one of the authors manually checked the URLs filtered to ensure that the URLs were effectively homepages: 159 unique URLs representing 10,987 visits (only 27 visits were wrongly identified as homepages). See details in Appendix S2. The rest of the news URLs (N = 53,228; 19,353 for wave 1 and 33,848 for wave 2) were scraped. Of those, 36,583 corresponded to news articles (see Data filtering), and we excluded 1,403 cases (resulting in 35,180) in which the difference between the user visit (in-situ) and the scraping (ex-situ) time was larger than twelve hours. Due to the characteristics of the browser extension, they occur when the individual leaves the browser tab open for a long time. Finally, we addressed page refreshes, which are reloads of the same page by the same user within a short period, e.g., due to loading issues. We first identify a time threshold that minimizes the discrepancies measured to provide reliable lower-bound estimates (see Appendix S16 for details). Specifically, a page refresh is defined as visits to the same URL by the same participant within 20 seconds, retaining only the last refresh. In total, 34,108 news articles were included in this study.

**Identification of news articles**

We broadly define a news article as a written piece that informs about any area of public interest and is published on a news website. Although a narrower definition of news articles, e.g., news of public interest or "hard news" (Reinemann et al., 2012), is more in line with scholarly tradition, successfully distinguishing such articles is intrinsically linked to the very error we are quantifying. Thus, we depart from a definition that is easy to operationalize through rules that exploit the URL structure for identifying news articles.

The identification of news articles relies on the observation that most news outlets follow recommendations to facilitate the indexing process of major search engines, called Search Engine Optimization (SEO). A well-known pattern in research (Clemm von Hohenberg et al., 2024) is to separate webpage titles via dashes (e.g., example.com/this-is-a-title). However, sometimes news articles are identified by a (relatively) long number (e.g., example.com/segment/2134324324) or a combination of both (example.com/the-title-3243234.html). We first exploit these patterns to identify this type of page (called leaf pages) in the structure of a website: the previous examples contrast with intermediary URLs such as example.com/politics/.

Once we have identified the leaf pages, we used keywords in the path of the URL to further exclude streaming content (e.g., podcasts or videos, including TV series, films, or talk shows), news live streams that offer updates on current events developments (e.g., the Israel-Hamas conflict, or sports events), only comment or forum pages and informative dashboards (e.g., weather, stock market figures, sports results). The list of keywords and the method for their selection (also based on the URL) is described in Appendix S3.

For the evaluation, one of the authors manually annotated an (imperfect) stratified sample (N = 1,005). The stratification was based on news domains and the two automatically annotated classes (article vs non-article). The goal was to sample 5 URLs per domain per class. Given the scarcity of URLs for some domains and the imbalance between the annotated classes (37,049 articles, 16,790 non-articles), it was impossible to achieve the exact quotas. However, we attained a ratio of 55% news articles to 45% non-articles, including at least one article from each domain while balancing the most visited domains. The method to identify news articles resulted in an F1-score of .937 (precision = .927, recall = .948). To better estimate the performance within the full tracked sample, we weighted according to domain frequency, resulting in an F1-score of .957 (precision = .932, recall = .984) (see Appendix S4).

**News content extraction**

The content of news articles gathered via in-situ and ex-situ is stored in plain HTML, which structures the text so browsers can arrange the elements on the user screen. HTML embeds a large amount of metadata, including scripts (javascript), visual information (CSS, Cascade Style Sheets), and data unique to the user session. Although extracting the raw text of a webpage is a deterministic process (as the purpose of HTML is precisely to separate the text from the structure), the extracted text will include boilerplate content from elements such as advertising, menus, sidebars, and footers. In contrast, the library trafilatura extracts the substantive content of news articles with high accuracy (Barbaresi, 2021; Bevendorff et al., 2023). We analyzed our results using the (raw) HTML, raw text extracted via the library selectolax (Artem Golubin, 2023), and cleaned text using trafilatura. To clarify, trafilatura is both a scraping and a content-extracting library, and we used it for both purposes in the present paper (see Data collection).

**News categorization**

We classify the news article using three strategies to show the discrepancies in in-situ vs ex-situ collection. The strategies exploit three levels of granularity of browsing data:
- **Domain-based:** Stier et al. (2020) offer a classification based on approaches to journalistic standards and content focus, concretely, commercial broadcasting, digital-born outlets, hyperpartisan news outlets, legacy press, public broadcasting, tabloid press, and portal news, e.g., via mail providers like MSN. The authors provide a list of domains (N = 679) mapped to each category.
- **URL-based:** A categorization assigned by the news outlet is often embedded in the path of the URLs, e.g., the URL "sample.com/national/politics/this-article" indicates an article related to national content and politics. For example, previous research has exploited the URL structure to build datasets for training automatic classifiers (Flaxman et al., 2016; Guess, 2021; Jürgens & Stark, 2022; Stier et al., 2022). We use this to divide the news articles into 16 non-mutually exclusive categories. The exact keywords used, class distribution, and the procedure are explained in Appendix S5. The method classifies 89.04% of the URLs, 75.46% excluding the generic category (indicating main news within the news outlets but no specific category).

- **Content-based:** We use a recently published ML classifier trained on press releases of political parties (Erfort et al., 2023) to categorize the articles according to their cleaned text (i.e., the one extracted with trafilatura). This classifier suits our case because news outlets often use these press releases for their reports and whose agenda is aligned closely with political elites regarding major political topics (McCombs & Valenzuela, 2020); we address shortcomings in the limitations section. The ML classifier includes a confidence score between 0 and 1 that can be used to filter results using a threshold score.

**Levenshtein distance**

Our dependent variable to answer our research questions is the normalized Levenshtein distance between the content gathered in-situ and the content gathered ex-situ. The Levenshtein distance measures the minimum number of single-character edits (insertions, deletions, or substitutions) required to change one text into another. It is normalized by dividing the number of characters of the longest word. Thus, the normalized distance can be interpreted as the percentage of change between two texts. For example, if the text extraction from an HTML of the ex-situ collection led to an empty string, the distance is 1 (as long as there is some text in the in-situ collection). If a scraping error results in a missing HTML, we treat it as a special case in which the text is also an empty string.

The Levenshtein distance is an adequate methodological choice, given that the expected changes in news articles are unaffected by the method's inability to capture semantic similarities. For example, two texts could be semantically equivalent and have a distance of 1 due to the possibility of writing the same message in entirely different ways. We argue that such major rephrasings are unlikely in published news articles. Moreover, it is theoretically impossible that scraping would cause such changes.

**Statistics**

We use linear regression and ANOVA to estimate the effects of our main factors (waves, libraries, post-hoc collections, and news categories) on the Levensthein distance of the content representations (cleaned text, raw text, or HTML). We estimate marginal means with confidence intervals and pairwise comparisons (TukeyHSD) using the emmeans package (Lenth, 2023). Time-distance associations to Levenshtein distances were evaluated with Pearson's correlations, and chi-square tests compared news category distributions pre- and post-debiasing. Unless otherwise specified, CI indicates 95% confidence intervals.

# Results

**Disparities between in-situ and ex-situ collections**

We investigated the discrepancies between the web content collected in-situ and the web content collected ex-situ (RQ1). We calculated the normalized Levenshtein distance between the two conditions in their HTML, raw text, and cleaned text as dependent variables. We find the estimated marginal means of the distances are higher than 0 for all representations of the content (H1a): $M$ = 40.1% (CI [39.8, 40.5]) for cleaned text, $M$ = 49.9% (CI [49.5, 50.3]) for raw text, and 71.8% (CI [71.5, 72.2]) for HTML across the two waves; statistically different from each other in the predicted directions (Tukey HSD): H1b: cleaned text < raw text (-9.73 percentage points; $p < .0001$) and H1c: raw text < HTML (-21.97 percentage points; $p < .0001$); see Appendix S7. The raw and cleaned text results distances are displayed in Figures 1A and 1B for waves 1 and 2, respectively. Given that the cleaning text library removes unnecessary components (e.g., menus) and highly volatile text that is generated dynamically (e.g., ads), cleaned text (Figure 1D) is skewed towards 0s and 1s; in general, we argue that it is fairer metric to analyze the data. For the rest of the paper, we will focus on the distance of the cleaned text unless otherwise specified.

Figure 1D supports our decision to exclude dynamic loading (including simulated scrolling) tools in our study (see Data collection); if such tools had a substantial potential to affect our results, then we would have observed a large number of cases in the middle distance values of the figure because they signify that the content is incomplete (e.g., due to partial loading or lack of scrolling). To verify this, we stratified sampled 30 cases with distances between .2 and .8 (i.e., 2 cases for distances between .2 and .3, and so on; 6 bins in total) and only found 3 instances in which dynamic loading might have helped. More generally, upon searching the keywords "javascript" and "js" on the cleaned text targeting messages such as "activate javascript"; only 487 URLs out of 34108 were found, mostly from outlets with subscription-based models (N > 406). Another 156 cases contain the keywords in the raw text with distances > .2.

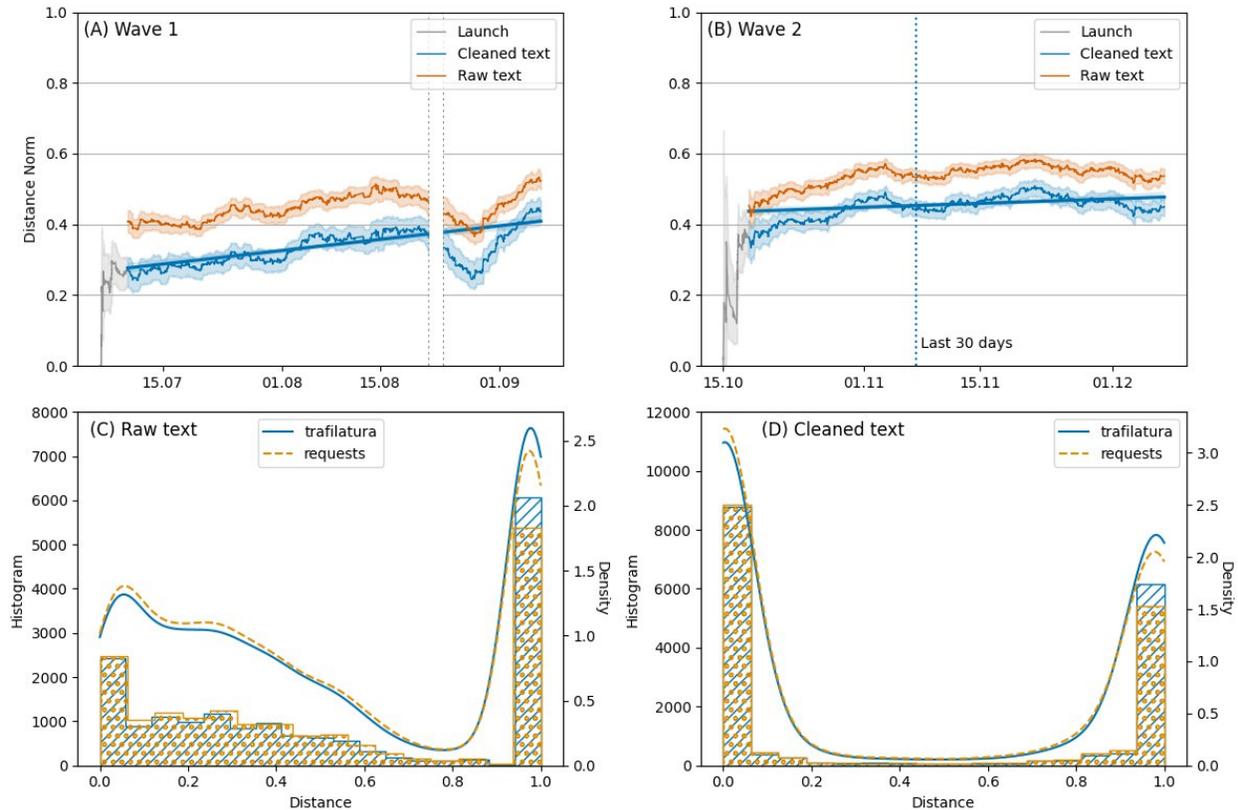

**Figure 1. Normalized Levensthein distances per wave, content type, and scraping library.** The top plots show the 7D rolled normalized average Levensthein distance between the near real-time ex-situ data collection and the in-situ data collection for the first wave (plot A, N = 12,326; from July 6th to September 6th, 2023) and second wave (plot B, N = 21,782; from October 15th to December 7th, 2023). In each plot, the two lines show the cleaned text (extracted using the trafilatura library, thin blue line at the bottom) and raw text (extracted using the selectolax library, thicker orange line at the top). Bands represent 99% confidence intervals. The straight lines correspond to linear fits between time and distance, excluding the gray areas on the left of plots, i.e., the initial launch period corresponding to the first three days of data collection in which the number of participants was low (and the server was not affected by heavy previous scraping, the gray line before the thin dark blue). The interruption in plot A is due to a technical issue during the collection between October 22nd and October 24th. The bottom plots show the distributions (histograms and density curves) per content type, raw (C) and clean (D), and split by scraping library for the near real-time ex-situ collection compared to the in-situ one. The legend indicates the color and line style; the diagonal filling of the bars represents trafilatura, and the circles represent the requests library.

Supporting (H1d) and (H1e), we found statistical differences between wave 1 (*M* = 33.8%, CI [33.0, 34.6]) and wave 2 (*M* = 45.4%, CI [44.8, 46.0]), p < .0001, and between the request scraper (*M* = 38.6%, CI [37.9,39.3]) and the trafilatura scraper (*M* = 40.6%, CI [39.9, 41.3]), p < .0001. The differences between the libraries are, however, only noticeable on wave 2 (TukeyHSD): -3.01 percentage points; p < .0001. An in-depth exploration revealed that spiegel.de blocked the trafilatura scraper between the waves. Therefore, we ran a model controlling for this (Appendix S9), no longer resulting in statistical differences between the scraping libraries except for wave 2 for spiegel.de (Tukey HSD, 74.24%, p < .0001); the statistical differences between the waves hold in all cases. A closer look at the time series in Figure 1 (i.e., 1A and 1B) reveals that the deterioration has a stronger association for wave 1

($r$(12,323) = .094, p < .0001) compared to wave 2 ($r$(21,779) = .03, p < .0001). The correlation is no longer significant for the last 30 days, i.e., the period used for the post-hoc experiment. See Appendix S10.

Regarding the progression of the discrepancies over time in the post-hoc experiment (RQ2), we can confirm H2: the more we postpone the ex-situ scraping, the larger the distance between the two collection methods (p < .0001), as in Figure 2A. The estimated marginal means of the distances for 0, 30, 60, and 90 days delay follow an incremental order: 46.6%, 52.1%, 52.9%, and 53.1%, i.e., the total increment in distance between the 0 and 90 days is 6.5 percentage points (p < 0.0001). However, the effect is concentrated in the first 30 days, i.e., the difference between 30 days and 0-day delay is 5.5 percentage points (p < .0001), while the difference between 30 and 90 days is .98 percentage points (n.s.). In a model that excludes the 0-day delay level (using the 30/60/90 delays as numerical), a weak trend in the predicted direction is still noticeable (p = .089); the differences between the scraping libraries hold over time even after removing spiegel.de (p = .009). See Appendix S11.

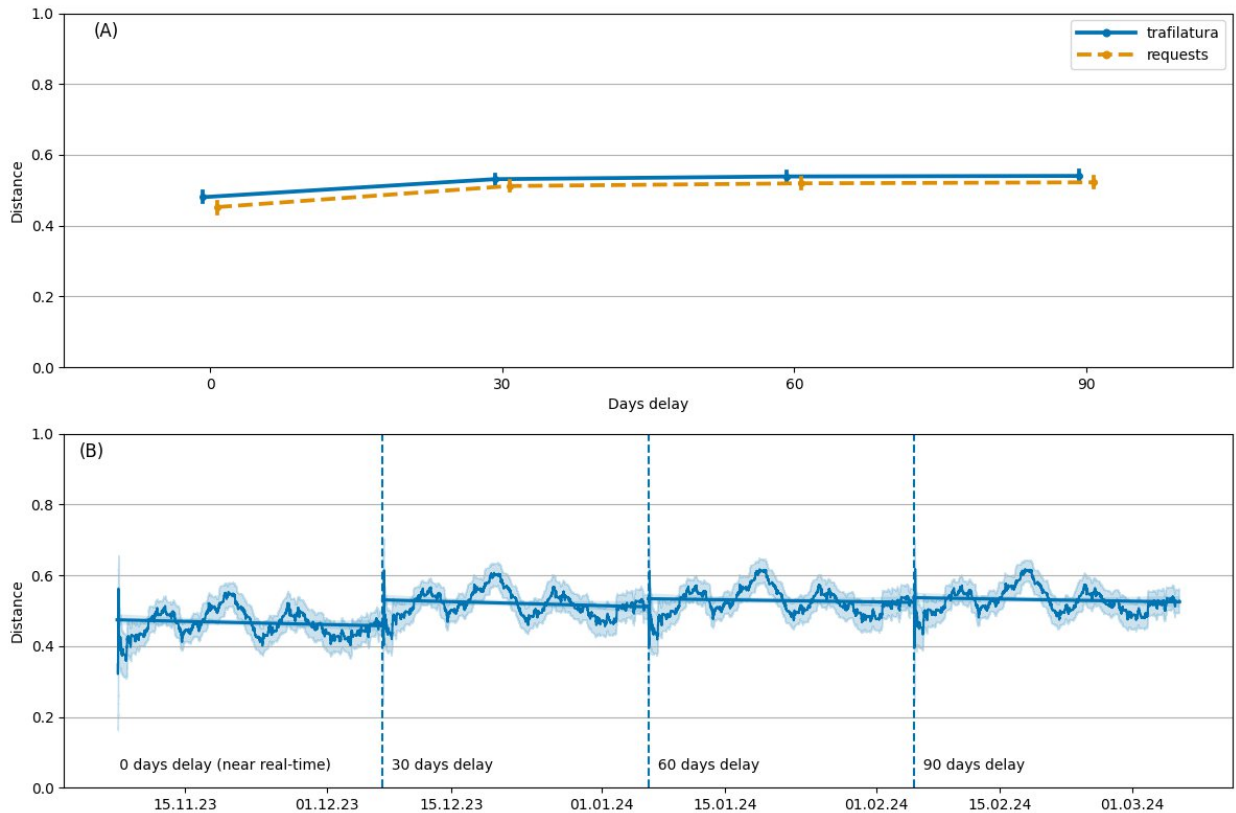

**Figure 2. Post-hoc ex-situ scraping experiment.** Plot (A) shows the average distances per library for each of the (post-hoc) ex-situ collections: 0 days (i.e., near real-time) and 30/60/90 days following user interaction compared to the in-situ collection. The vertical lines on top of the averages represent 99% confidence intervals. Plot (B) shows a time series using 7D rolled average distances for the same period. The first section (0 days) corresponds to the last 30 days of wave 2 (Figure 1B). Bands represent 99% confidence intervals. The straight, continuous lines represent a linear fit between time and distance. The vertical dashed lines indicate the end of one post-hoc collection and the beginning of the next.

**Disparities across news types**

We investigated the distances in the cleaned text across different categorizations of news (RQ3). Supporting H3a, H3b, and H3c, respectively, we find statistical differences for the news categorization based on (a) the news type list provided by Stier et al. (2020) (p < .0001), (b) the keywords applied to the URLs (p < .0001) and the ML classifier (Erfort et al., 2023) (p < .0001). For the latter, the effect increases (i.e., higher F values) as we increase the accuracy of the classifiers (via a score threshold). See Appendix S12. Figure 3 shows the distances according to each category.

We also included the clean and raw text in Figures 3A and 3B to observe the effect of the removed boilerplate; there are clear differences depending on the domain base categorization (Figure 3A). We also observe (Figure 3A) that the distances between the raw text and the cleaned text for different types of news vary, in particular, according to the domain-based categorization; for example, the difference for tabloids is 29.56% (CI [28.90, 30.21]) while for commercial broadcasters is 1.96% (CI [.63, 3.29]). For other means and contrasts, see Appendix S13.

We further explored the distances per domain (Appendix S14) and found that msn.com, which comprises 31.3% of the visits (N = 10,690), is responsible for the high value for the distances of portal news (Figure 3A).

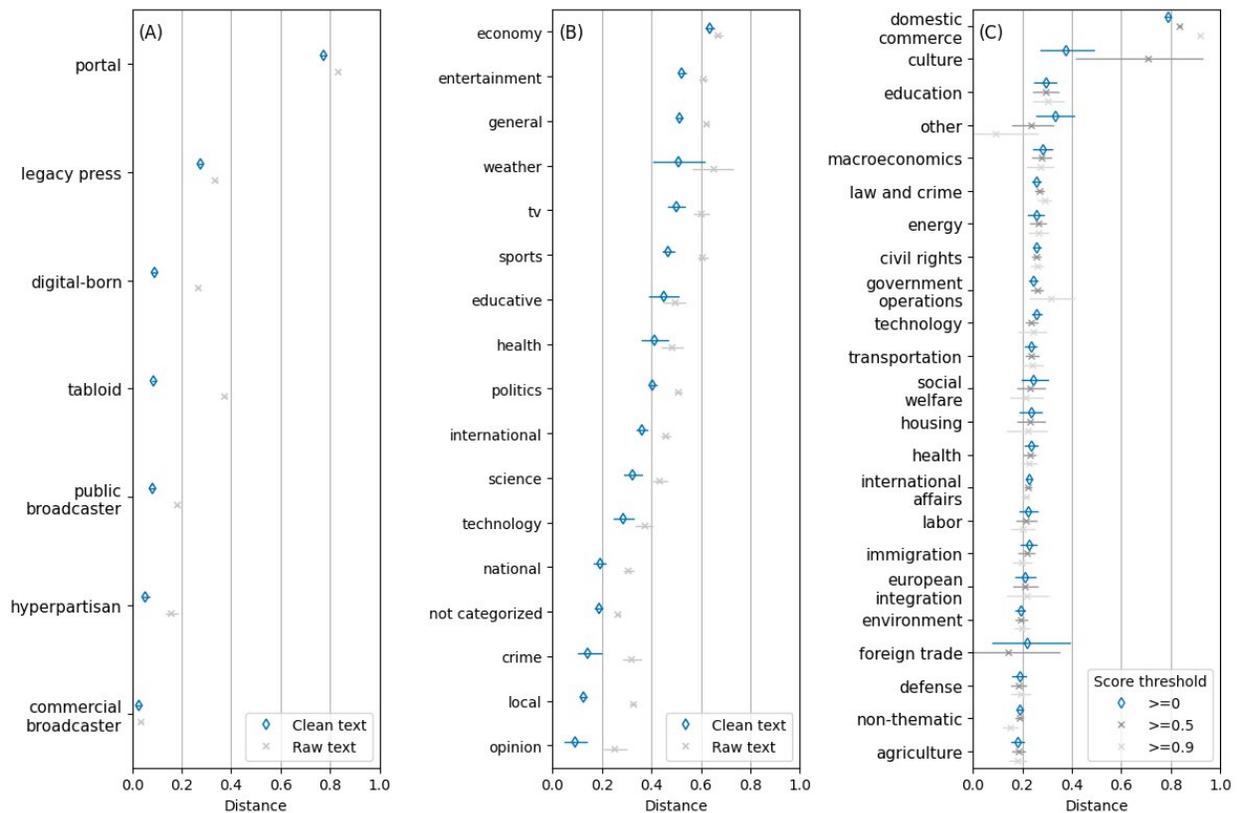

**Figure 3. Normalized Levensthein distances for different news classifications.** The left plot (A)

shows the average distances for each news type based on their domains. The center plot (B) shows the mean distances for each news type based on the URL paths (one URL can belong to several categories). The legends for plots (A) and (B) indicate the type of text used to calculate the distances. The right plot (C) shows the mean distances for each news type based on their content. The legend indicates the threshold used to filter the category; the higher the threshold, the higher the confidence that the article belongs to the category. The lines on top of the means represent 95% confidence intervals.

**Mitigating biases**

We further examine the biases corresponding to Figure 3C by inspecting the number of articles identified per category in the in-situ vs ex-situ collection for different score thresholds of the ML classifier (Figure 4, top row). We highlight the following observations:
> (1) the *Domestic Commerce* category contained 78.2% more articles for the **in-situ** collection for a threshold of 0 (Figure 4A) compared to the ex-situ collection, 83.8% for a threshold of .5 (Figure 4B), and 89.1% for .75 (Figure 4C).
> (2) conversely, the *Non-thematic* category contained 385.7% more articles for the **ex-situ** collection for a threshold of 0 (Figure 4A), and
> (3) the *Technology* category contained 31.08% more articles for the **ex-situ** collection for a threshold of .5 (Figure 4B).

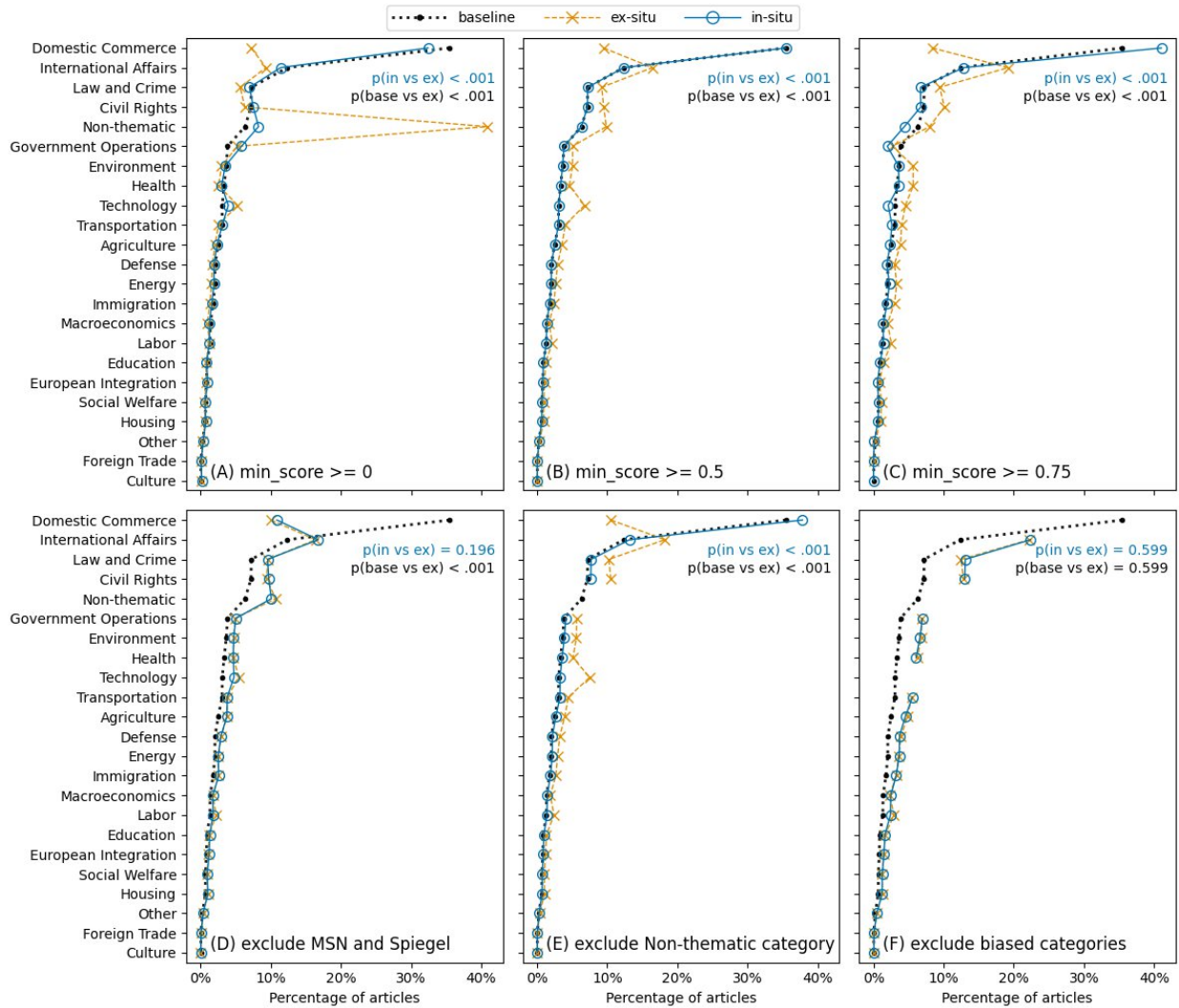

**Figure 4. Strategies to counteract systematic biases in the distributions.** The plots show the distribution of the articles (X-axis, percentage) across content categories (Y-axis) for the two collection methods, ex-situ and in-situ, using different debiasing strategies. The top row shows the distribution using three score thresholds on the automatic classification: (A) no threshold, (B) .5, and (C) .75. The bottom row corresponds to three different strategies: (D) removing the two most problematic domains, (E) excluding the Non-thematic content category and (E) removing the biased categories: Non-thematic, Domestic Commerce, and Technology. The black dotted line in all plots is used as a baseline and corresponds to the in-situ with a score threshold of .5 (i.e., in-situ distribution in plot B). The p-values in the top-right correspond to chi-square tests for independence of the in-situ vs. ex-situ collection method with the filters of the plot (blue) and the baseline vs. ex-situ (black).

We further inspected the data. First, 77.5% of pages classified as *Domestic Commerce* in the in-situ collection are, instead, classified as *Non-thematic* in the ex-situ collection, with a low score ($M$ = .15, $SD$ = .05), explaining observations (1) and (2) for a threshold of 0 (Figure 4A). This stands to reason: the *Non-thematic* category is inflated by errors in the ex-situ collections (mostly coming from *Domestic Commerce*). Second, we found that 1.14% of pages classified as *Domestic Commerce* for the in-situ collection are classified as *Technology* in the ex-situ collection with a high score (*M* = .71, *SD* = .2), shedding initial light in observations (1) and (3) for the .5 threshold (Figure 4B). To further understand why, we inspected the top 100 most miss-categorized articles (according to the distance in the scores): 50 involving *Domestic Commerce* and 50 involving *Technology* (See Appendix S15 for details). 84% corresponded to pages that required user interaction, such as paywalls and logins (73%); presumably, they were wrongly classified due to the presence of commercially-related words (e.g. "payment" and "subscription") and technical words (such as "username" and "password").

We attempted several filters to de-bias the collection; specifically, the goal is to obtain a similar distribution of categories in the in-situ and ex-situ collections (while reporting excluding criteria). First, increasing the score of the automatic classifier does not address the bias (Figure 4A to 4C). Second, removing the most problematic domains highlighted in previous sections (i.e., msn.com and spiegel.com) does not address the bias (Figure 4D). Third, removing the *Non-thematic* category does not remove the bias (Figure 4E). Fourth, only removing *Domestic Commerce*, *Technology,* and *Non-thematic* categories aligns the two distributions (Figure 4F); $\chi^2(19, N = 28,764) = 16.87, p = .599$. We will discuss the implications of these results.

## Discussion

Due to the relative stability of online news article content compared to other web resources (e.g., homepages and news listings), researchers might be inclined to believe that scraping news articles from web browsing histories is less susceptible to bias. This assumption rests on the premise that potential errors introduced during the scraping process would be randomly distributed and have minimal impact on the overall conclusions. We demonstrated that this is not necessarily the case. The errors can under-represent particular news outlets, affecting the distribution of different news categorizations. It could be either because particular news outlets specialize in certain dimensions of news content or because a subgroup of users with corresponding preferences concentrates their consumption on that specific outlet.

We also demonstrated that the discrepancies introduced by the collection environment (up to 46.6% between the in-situ vs. near real-time ex-situ in the post-hoc experiment, Appendix S11) are orders of magnitude higher than those added by the delay of the scraping process by several months (6.5 percentage points). However, it is noteworthy that most of the post-hoc discrepancies occurred during the first month; it strongly indicates that the relation is not linear, which should be considered when interpreting other works. For example, Dahlke et al. (2023) took their first measurement one year after the user interaction; their last measurement (one year after) differed by only 2.2 percentage points (for "hard news"). Furthermore, our experimental design identified a specific mechanism in which the scraping process deteriorates over time: one of the two scrapers, i.e., trafilatura, was blocked by a major news domain. We attributed this to the consistent use of one user agent by trafilatura while implementing a randomization process for the requests library (see Data collection). More outlets will likely block the scrapers as requests accumulate over time. As additional evidence outside the post-hoc experiment, we saw increasing discrepancies in waves 1 and 2: the association in wave 1 was stronger because there were fewer participants, thus delaying the deterioration, or because wave 2 started one month after the end of wave 1, hence suffering a carry-over effect. Either case points to a worsening performance due to an excess of requests. Hence, the estimated marginal mean of wave 1 (33.8%) serves as a lower-bounded estimate for static scraping approaches of low intensity per server. Crucially, the carry-over effect was not significant for the post-hoc experiment, as we only selected the last 30 days of wave 2 and we started the experiment immediately after wave 2; therefore, all post-hoc collections would have suffered from previous weeks of scraping from hundreds of users; thus, 46.6% is a better lower-bounded estimate for a more intensive and prolonged scraping process.

We found biases across the three content categorizations we tested: domain, URLs, and content. The interests behind the news outlets can explain the differences in the news categorized by domain. Commercial broadcasters and hyperpartisan news outlets displayed the least differences; the accessibility of their content (via the ex-situ method) could be explained by the underlying revenue strategies behind these providers: commercial broadcasts finance their news primarily via advertising, and hyperpartisan media have a political agenda that incentivize removing barriers to access their content. Conversely, the content of the legacy press is more often blocked behind paywalls and, thus, more challenging to scrape; this aligns with previous findings indicating that "hard news" is more affected by temporal effects than, e.g., misinformation web pages (Dahlke et al., 2023), as "hard news" are more likely to be behind paywalls (Myllylahti, 2017). Regarding financial strategies, the discrepancies in the raw text were more pronounced for, e.g., tabloid news, suggesting an overload of peripheral content, such as advertising and internal referrals, that characterize these news outlets. The other news classifications, by URLs and by content, complement each other: the classification by URLs is intrinsic, i.e., provided by the news outlets themselves, while the classification by content is extrinsic, i.e., provided by an ML classifier trained in a different collection. The biases found in the URL and content classifications may be simply a byproduct of the non-uniform distribution of the error within certain news domains (i.e., the domain as a mediator). Alternatively, news outlets might be compelled to facilitate access to certain content (e.g., regional content) while putting others (e.g., economy, entertainment, TV) behind paywalls as individuals might be more inclined to pay for them (e.g., Myllylahti, 2017; Sjøvaag, 2016).

In our exploration of bias mitigation, we evaluated three strategies. Neither the attempt to increase the accuracy of the ML classification (via higher score thresholds) nor removing the identified problematic domains addressed the bias. The only successful strategy was to remove the biased categories, with the caveat that we were able to identify them because we could access the precise distribution via the in-situ collection. Nevertheless, our examination of these categories revealed that most biases emerged due to web components requiring user input, such as paywalls, logins, or cookie consents; this aligns with the skewed distribution towards zeros and ones of the cleaned text, which indicates that direct changes in the texts of the news articles only play a minor role, as opposed, to the full accessibility of the news piece, e.g., due to subscription mechanisms (paywalls, logins and, arguably, cookie consent). We emphasize the crucial need for a robust method to identify such content types and test such mitigation strategies further.

**Methodological implications**

Our study boasts a high level of ecological validity due to its reliance on individuals' actual web browsing behaviors rather than a simulated or systematically controlled data collection process. We argue that emulating the web browsing time in the scraping process (simulating traffic patterns of the participants) contributed to our low HTTP error rate as it spreads the requests of the ex-situ collections over a period of time equivalent to the in-situ collection. Additionally, we ensured that the delayed post-hoc collections would not overlap with the near real-time collection (i.e., we were never collecting data simultaneously) and excluded homepages (reducing the number of server requests), which should prevent systematic blockages by the servers, meaning that the measured discrepancies represent best-case scenarios. Meanwhile, we also canceled the disproportionately high success rate of the initial requests, as our calculations are based on the last 30 days of user interactions, and we vaunt about the precision and time-sensitivity of our approach as our post-hoc collection are calculated exactly after each user interaction's timestamp.

Using clean text extraction through trafilatura reduces noise and ambiguities in the data, leading to more evident results concentrated at the extremes (0s and 1s). This approach improves the clarity of our results as it filters out boilerplate content, such as menus, and dynamic or personalized content, like ads (which introduce noise by decreasing or increasing the text similarity).

**Recommendations for ex-situ data collections**

Although there are evident shortcomings in ex-situ data collection of web browsing content, we recognize the intricacies inherent to web-tracking collections that include the full content of the visits, which may limit its accessibility to all researchers. These intricacies vary from more elaborated technical setups (i.e., to support the secure uploading and management of HTML) to privacy concerns that lead to potentially lower participation rates (Makhortykh et al., 2021). Given these constraints, more superficial forms of web-tracking can still be profitable in various research scenarios. In particular, the analysis of URLs can yield considerable insights. Our research demonstrates this potential, assigning a category to 89.04% of articles using URL data alone; a recent study demonstrated that state-of-the-art natural language processing methods can achieve very good performance using URLs alone (Schelb et al., 2024). Thus, the use of the URLs and associated metadata (such as timestamps and visit durations) should be considered before delving into the content analysis.

However, the content behind a URL is a rich resource, and scraping is the most accessible method to (try to) gather it, and a robust infrastructure is crucial to minimize the discrepancies between the data observed by the participants and the data analyzed by the researchers. For example, our scraping procedure limited the request rate by spreading the requests over a generous time equivalent to the data collection instead of scraping all the URLs as fast as possible. We also incorporated the randomization of user agents for the request library. These simple strategies can increase the robustness of the scraping; however, more intricate techniques can improve the data collection further. Rotating IPs and repeated attempts due to failures (though note that HTTP errors were minimal in our case) could help distribute the number of requests and account for potential connectivity problems. The intensity of the scraping can be further reduced by focusing only on the relevant URLs for the study. Although we argued that dynamic content plays only a minor role in scraping news articles is still a generally more robust alternative than more commonly used static libraries. Additionally, tools like Selenium enable interactions with web elements, e.g., accepting cookie consents or dealing with buttons that must be pressed to load the full content. However, it comes with challenges: the need for simulated behaviors must be identified and programmed on a per-domain basis, which is beyond most researchers' resources. Finally, some commercial online services (i.e., APIs) offer some of these strategies; nevertheless, researchers should be careful about their use, as browsing histories can contain personal information of individuals, and their consent to re-share is necessary. Ideally, the academic community would pool resources (e.g., academic IPs) and efforts (e.g., scripts interacting with dynamic elements) to build secure and robust academic scraping infrastructures.

Researchers should develop effective methods to identify typical sources of errors, paywalls, human verification tests (e.g., CAPTCHAs), cookie consent pop-ups, login pages, and web elements that suggest the text is incomplete (e.g., some news outlets require to press a button to display the full article). Identifying these elements allows for several approaches to address biases: (1) removing such pages from the analyses, (2) running robustness tests, e.g., a complementary analysis that introduces weights proportional to how much domains are affected, (3) identifying and transparently report and/or remove domains that are not suitable for analysis so that results can be interpreted in light of the limitations, and (4) seeking specific methods to increase the coverage for certain domains. For the latter, researchers can use other repositories, such as the Internet Archive and specialized news databases, enter into direct collaborations with news outlets, or implement specific solutions for news outlets that require user interactions to display the full content (e.g., accepting cookie consents or pressing a "load full article button"). Nevertheless, researchers should also be mindful that the participants might have never seen the content because the same prompts deterred them. Therefore, these measures should be accompanied by a mechanism to discard short visits that might signal failed attempts from the user side.

**Limitations**

The sample is restricted to Germany. While the results illustrate biases likely to emerge in any other country, the news environments differ, and the distributions displayed in the article are only relevant to it. An incentive experiment was conducted during the data collection, which might have influenced their activity patterns and the number of news domains visited. The sample is nonrepresentative and only includes desktops; despite capturing authentic user behaviors, the discrepancies limit the generalizability of our results. However, (1) this deviation is a common challenge in web-tracking studies where participant recruitment can be complex, (2) our estimates include two waves with different participants who are partly recruited from a probability-based survey, (3) although the focus on desktop usage has been criticized (Reiss,

2023; Yang et al., 2020), web-tracking on other devices would similarly be affected if the content is scraped, and (4) should the biases reported in this paper be exclusive to certain subgroups represented in our sample, the resulting biases are aggravated even further from a substantive point of view.

The classification of articles using URL paths depends on news outlets' internal categorization, which varies and affects reliability. Nevertheless, the approach has been used to train content classifiers, and in our case, it has displayed good coverage. Additionally, we remark that we only used it to illustrate the biases. Similarly, using a pre-trained machine learning classifier (developed training on political press releases) would affect performance, given that news articles do not perfectly fit the party press range of topics. However, as this classifier is uniformly applied to the same content (i.e., in-situ collection) for both in-situ and ex-situ datasets, any systematic errors are expected to equally impact both, preserving the validity of our comparative analysis of the distribution patterns across these datasets. Additionally, we evaluated different score thresholds and demonstrated that the found systematic biases can be addressed by targeting issues related to the ex-situ collection.

Since we randomly allocated the URLs to each library to keep the scraping low intensity (see Data collection), our results relied on only two libraries, as adding more would have reduced our degrees of freedom for the statistical tests. However, the two libraries are widely used, and we only found small statistical differences between their performance. We attributed such differences to their default configurations of the HTTP parameters, e.g., the impact of the randomization of user agents. Other parameters are less likely to affect the scraping process unless impractical values are selected, e.g., request expiration timeouts lower than a few seconds.

Our data collection scraping procedure is static, as the selected libraries cannot capture dynamically loaded content. Nevertheless, we argued and presented supporting results that static scraping is suitable as we focused exclusively on news articles often loaded entirely in one request (instead of dynamic platforms, which are more common in, e.g., social media feeds). In our case, running the experiments in near real-time added to the complexities, including running and monitoring heavy processes. We emphasize that logins and paywalls were responsible for 73% of the discrepancies analyzed in our exploration of the biases, for which account registration is also necessary.

## Conclusion

Over the past decades, free access to online content has declined (Arrese, 2016; Bruns, 2021; Freelon, 2018; Myllylahti, 2017; Sjøvaag, 2016). Consequently, web scraping has emerged as a tool for researchers to gather online data (Freelon, 2018). However, our study indicates that web scraping may not provide the high-quality data needed to measure media exposure and its effects accurately, even in a domain outside of the known restrictions of social media platforms, such as news outlets. We further emphasize the urgency to continue developing robust in-situ data collection methodologies and democratize their access while, at the same time, improving methods to mitigate the biases introduced by web-scraping approaches.